\documentclass[12pt,amsmath,amssymb,aps,preprint]{revtex4}
\usepackage{graphicx}
\usepackage{amssymb}
\usepackage{subfigure}
\usepackage{psfrag} 
\newcommand{\be}{\begin{equation}}
\newcommand{\ee}{\end{equation}}
\newcommand{\bel}[1]{\begin{equation}\label{#1}}
\newcommand{\bea}{\begin{eqnarray}}
\newcommand{\eea}{\end{eqnarray}}
\newcommand{\ba}{\begin{array}}
\newcommand{\ea}{\end{array}}

\newcommand{\exval}[1]{\mbox{$\langle \, {#1}\, \rangle$}}

\renewcommand{\eta}{n}

\begin{document}

\title{Hydrodynamics of the zero-range process in the condensation
regime} \author{G.M. Sch\"utz\footnote{e-mail:
    g.schuetz@fz-juelich.de} and R.J. Harris\footnote{Present address: Fachrichtung Theoretische Physik, Universit\"at des Saarlandes, 66041 Saarbr\"ucken, 
Germany; e-mail: harris@lusi.uni-sb.de}}


\affiliation{Institut f\"{u}r Festk\"{o}rperforschung,
Forschungszentrum J\"{u}lich, 52425 J\"{u}lich, Germany.}

\date{December 7, 2006}

\begin{abstract}
We argue that the coarse-grained dynamics of the zero-range process in
the condensation regime can be described by an extension of the
standard hydrodynamic equation obtained from Eulerian scaling even
though the system is not locally stationary.  Our result is supported
by Monte Carlo simulations.

\vspace{\baselineskip}

\end{abstract}

\maketitle

\section{Introduction}

The zero-range process (ZRP) was introduced in 1970 by Spitzer
\cite{Spit70} as a system of interacting random walks, where each
lattice site $k$ is occupied by $n_k$ particles which hop randomly to
other sites. The hopping rates $w_n$ depend only on the number of
particles $n$ at the departure site. Under certain conditions on the
rates $w_n$ and the particle density (see below) the grand-canonical
stationary distribution is a product measure, i.e. there are no
correlations between different sites \cite{Andj82}.  An exact
large-scale description of the {\it dynamics} has been proved for
arbitrary initial densities in terms of a hydrodynamic equation for
the coarse grained particle density $\rho(x,t')$, provided the rates
are non-decreasing, i.e., $w_{n+1} \geq w_n \quad \forall \quad n$
\cite{Reza91,Kipn99}. In this so-called attractive case the density
satifies the continuity equation 
\bel{1-1}
\partial_{t'} \rho + \partial_x j(\rho) = 0 
\ee 
where $j(\rho)$ is the
stationary current in the grand-canonical distribution with density
$\rho$.

Depending on the choice of rates, in non-attractive systems a rich and
rather varied dynamical and stationary behaviour emerges, for a recent
review see \cite{Evan05}. In particular, the model may admit a
condensation phenomenon analogous to Bose-Einstein condensation. In a
periodic chain one then finds that, above a critical density $\rho_c$,
a finite fraction of all particles in the system accumulate at a
randomly selected site, whereas all other sites have an average
density $\rho_c$ \cite{Oloa98,Jeon00a,Jeon00b,Gros03a}. The late-time
dynamics of condensation has been studied in terms of a coarsening
process \cite{Gros03a,Godr03}. For analysis of the early time range
Kaupu{\v z}s et al.~\cite{Kaup05} have generalized an equivalent
exclusion process for traffic flow \cite{Anta00}. They observed a
metastable regime as a precursor of the coarsening process.  The
metastable state with a current above the critical current persists
for some finite time range which depends on the hopping rates.

If valid, the hydrodynamic description (\ref{1-1}) would apply to
times much later than this metastable regime.  Interestingly, however,
the proofs for the hydrodynamic limit do not work when the condition
on the rates $w_n$ for condensation is met. The reason for this
failure is {\it not} a minor technical issue of the standard
hydrodynamic approach but lack of local stationarity during the
coarsening process, the early stages of which fall within the
hydrodynamic time regime. This profound violation of one of the most
basic assumptions of hydrodynamic theory in conjunction with the
highly discontinuous condensation phenomenon may lead one to suspect
that (\ref{1-1}) might not be valid for initial density profiles where
some region of space is above the critical density $\rho_c$. It is the
aim of this paper to show that such a view, even though
well-motivated, is overly pessimistic. We argue that, properly
interpreted, the hydrodynamic limit (\ref{1-1}) is robust and valid
also in the condensation regime. The theoretical analysis of
Sec.~\ref{s:theory} is supported by Monte Carlo simulation in
Sec.~\ref{s:sim}.

\section{Totally asymmetric zero-range process}
\label{s:model}
To set the stage for the new ideas in the next section, we now define
the details of our model and review some known results including the
hydrodynamic limit for subcritical densities.

For definiteness we consider here the one-dimensional totally
asymmetric zero-range process (TAZRP) with periodic boundary
conditions. The one-dimensional ZRP has attracted particular interest
since it can be mapped to an exclusion processes for which
double-occupancy of sites is forbidden. The ZRP-particles are turned
into particle clusters between consecutive vacant sites (which
correspond to the sites on which the ZRP is defined). Condensation
thus corresponds to phase separation between a macroscopic particle
cluster and a disordered domain which also contains vacant sites. The
$n$-dependence of the hopping rates corresponds to a length-dependent
rate of detachment of a particle from a cluster of length $n$. The ZRP
has thus served for deriving a quantitative criterion for the
existence of non-equilibrium phase separation \cite{Kafr02} in the
otherwise not yet well-understood driven diffusive systems with two
conservation laws~\cite{Toth03,Schu03}.  Alternatively one can map
ZRP-particles onto strings of vacant sites between consecutive
particles (which correspond to the sites on which the ZRP is
defined). In this mapping the $n$-dependence of the hopping rates
$w_n$ translates into a distance-dependent hopping rate for exclusion
particles as may be expected for one-dimensional driven motion which
is embedded in three-dimensional space. This could be of importance
not only for the understanding of vehicular traffic, but also for
obtaining insight into the dynamics of ribosomes along m-RNA
\cite{Schu97} or the motion of molecular motors along microtubuli
\cite{Nish05}.  Other fields of application of the one-dimensional ZRP
include experiments on condensation and metastability in granular
media \cite{Lipo02,Shim04}.

If $\rho \leq \rho_c$ then the grand-canonical stationary product
measure has one-site marginals $P^\ast(n)=\text{Prob}[n_k=n]$ given by
\bel{1-2} 
P^\ast(n) = \frac{1}{Z} \phi^n \prod_{i=1}^n w_i^{-1}.  
\ee
Here the empty product ($n=0$) is defined to be 1, 
\bel{1-3}
Z=\sum_{n=0}^\infty \phi^n \prod_{i=1}^n w_i^{-1} 
\ee 
is the local ``partition function'', and $\phi$ is the fugacity which
determines the density $\rho=\phi (d/d\phi) \ln{Z(\phi)}$. Due to
particle conservation the product distribution defined by (\ref{1-2})
is stationary for every $\phi$ for which $Z$ exists. In the TAZRP
particles hop with rate $w_n$ from site $k$ to site $k+1$ on a
periodic chain with $L$ sites. This process satisfies pairwise balance
\cite{Schu96}, leading to a macroscopic stationary current
\bel{1-4} j
= \sum_{n=1}^\infty w_n P^\ast(n) = \phi.  
\ee 
The convexity of $Z$ ensures that the current is an increasing
function of the density.  In the case of condensation the radius
$\phi_c$ of convergence of the partition function is finite, with a
critical density $\rho_c < \infty$ as $\phi$ approaches $\phi_c$. The
product measure does not exist for densities beyond $\rho_c$.

An intuitively convenient starting point for a coarse grained
hydrodynamic description of the dynamics of the TAZRP is the lattice
continuity equation
\bel{1-6} 
\frac{d}{dt} \rho_k(t) = j_{k-1}(t) - j_k(t) 
\ee 
for the expected local density $\rho_k(t) =
\exval{n_k(t)}$, starting from some initial distribution. Here
\bel{1-7} 
j_k(t) = \sum_{n_k = 1}^\infty w_{n_k} P(n_k,t) 
\ee 
is the expected local current with the probability $P(n_k,t) =
\exval{\delta_{n_k(t),n_k}}$ of finding $n_k$ particles on site $k$ at
time $t$. We consider Eulerian scaling where the lattice constant $a$
(so far implicitly assumed to be unity) is taken to zero and the
system is studied for rescaled time $t'=ta$, i.e., the microscopic
time $t$ is taken to infinity such that the macroscopic time $t'$ is
fixed. Since we are working with a periodic chain with $L$ sites we
take $a=1/L$ and correspondingly $t=Lt'$. In the hydrodynamic limit
$L\to\infty$ the discrete chain of $L$ sites becomes a continuous ring
of circumference 1.

We first consider the subcritical regime where initially $\rho_k <
\rho_c$ everywhere on the lattice. In this case one obtains formally
$\partial_{t'} \rho(x,t') + \partial_x j(x,t') = 0$ by setting $k=xL$
in the lattice continuity equation and Taylor expanding in $1/L$.  In
order to arrive at the continuity equation (\ref{1-1}) one proves
local stationarity which can be done rigorously for the ZRP and other
lattice gas models under fairly generic circumstances.  Local
stationarity means that in the local environment of the point $x$, i.e.,
in a large but finite lattice region around the lattice point $x=kL$,
the system is found in its stationary state. Physically, it follows
from considering the limit where the microscopic time $t \to \infty$,
which allows all nonconserved (fast) local degrees of freedom to relax
to their local stationarity distribution at the local density $\rho$
(which because of the conservation law is a slow dynamical variable).
The identification of the expected local density $\rho(x,t')$
(appearing in (\ref{1-6})) with the coarse grained density of the ZRP
(appearing in (\ref{1-1})) comes from the law of large numbers and
local stationarity ensures that $j(x,t')$ is the stationary current
$j(\rho(x,t'))$ computed from the product measure; for details see
\cite{Kipn99,Spoh91}.

The hydrodynamic equation (\ref{1-1}) can be solved by writing
$\partial_{t'} \rho + \partial_\rho j(\rho) \partial_x \rho = 0$ and
using the method of characteristics. These are the lines
$x(t')=v_{\text{char}} t'$ along which the density remains
constant. The characteristic velocity is given by
$v_{\text{char}}(\rho) = \partial_\rho j$. One finds smooth segments
of the density which, depending on the initial data, may evolve into
shocks.  These are density discontinuities where the density jumps
from a value $\rho_\text{left}(x_s,t')$ to $\rho_\text{right}(x_s,t')$
at the shock position $x_s$. Shocks travel with velocity $v_s =
(j_\text{right}-j_\text{left})/(\rho_\text{right}-\rho_\text{left})$.
and are stable if the Lax condition $v_{\text{char}}(\rho_\text{left})
> v_s > v_{\text{char}}(\rho_\text{right})$ is satisfied.  An initial
density discontinuity which does not satisfy the stability condition
for shocks evolves into a rarefaction wave which is a smooth entropy
solution of the continuity equation (\ref{1-1}) \cite{Kipn99}.

On the microscopic scale, a shock is a sharp increase of the local density,
averaged over a finite lattice segment. The shock position has
diffusive fluctuations around its deterministic mean displacement
$x_s(t') - x_s(t'_0)= v_s (t'-t'_0)$ \cite{Spoh91}. The microscopic
objects corresponding to the characteristics are spatially localized
 finite perturbations of the local density which
travel with the collective velocity $v_{\text{coll}} =
v_{\text{\text{char}}}$ of the lattice gas \cite{Schu00}.  These microscopic
perturbations are analogous to kinematic waves \cite{Ligh55} appearing in
the
macroscopic PDE-description of nonequilibrium many body systems.
In view of this correspondence, we shall apply the intuitively appealing
term kinematic wave (in slight abuse of language) also to travelling
microscopic perturbations observable on the lattice scale.

In this way one can compute the macroscopic evolution of an initial
density profile provided that the initial density $\rho_0(x)$ is
subcritical everywhere. For hopping rates $w_n$ which do not lead to
condensation ($\rho_c = \infty$) this is not a
restriction. Furthermore, the open system behaves in a way analogous
to usual lattice gases with open boundaries
\cite{deMa84,Schu93,Derr93,Kolo98,Gros03b,Baha04}.  One can compute
the evolving density profile from (\ref{1-1}) even if the condition
for condensation is met, since the bulk density remains subcritical at
all times \cite{Levi05}.

\section{Hydrodynamics in the condensation regime}
\label{s:theory}
The standard considerations of the previous section fail for
$\rho_0(x) \geq \rho_c$.  For $\rho_0(x) > \rho_c$ the current cannot
be computed from the product measure and already at $\rho_0(x)=\rho_c$
the interpretation of the characteristics as local perturbations away
from the local density becomes open to doubt. A fluctuation below
$\rho_c$ would surely travel with collective velocity $v_{\text{coll}}
= v_{\text{char}}$, but the interpretation of a fluctuation above
$\rho_c$ becomes dubious. In order to argue that nevertheless the
continuity equation (\ref{1-1}), properly interpreted, describes the
macroscopic time evolution of the density under Eulerian scaling we
now consider a supercritical density segment $\rho_0(x) > \rho_c$.
The key observation is that local stationary is a sufficient, but not
a necessary condition for (\ref{1-1}) to be valid.

\begin{figure}
\centerline{\includegraphics[scale=0.8]{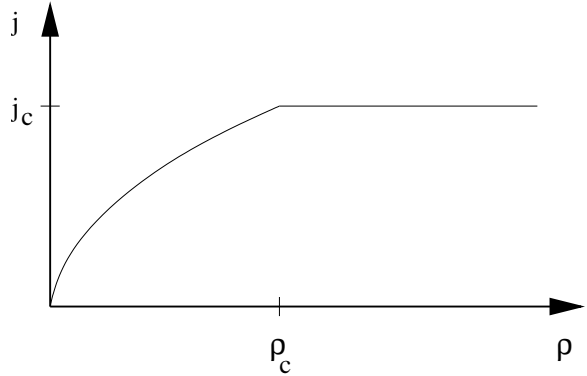}}
\caption{Schematic picture of the stationary current density relation
in the ZRP with condensation. Above the critical density the current
is constant with $j_c=\phi_c$.} \label{current}
\end{figure}

To fix ideas we first assume an initial distribution of particles such that
the whole chain has supercritical density $\rho_k > \rho_c$.  In a
finite system with $N$ particles the stationary current does not
increase beyond $\phi_c$ in the thermodynamic limit $L,N \to \infty$
with $\rho=N/L$ fixed even though finite-size corrections can be
substantial \cite{Evan05}.  Therefore the current-density relation
takes the form (Fig.~\ref{current}) 
\bel{1-8} 
j =
\begin{cases}
\phi & \text{for $0 \leq \rho \leq \rho_c$} \\ 
\phi_c & \text{for $\rho > \rho_c$}.
\end{cases}
\ee 
This behaviour can be rationalized by viewing the site $k_0$ where
the condensate is located in the infinite system as a source of
particles which are emitted at constant rate $w_\infty$ onto site
$k_0+1$. On the other hand, site $k_0$ acts as a sink which absorbs
all particles arriving from site $k_0-1$. The condensate site itself
contains always an infinite number of particles, defined such that for
every finite $L$ one has $N=\rho L$ particles in the chain. This
effectively breaks the ring into an open chain with a source at the
left boundary where particles are injected with rate $w_\infty$ and a
right boundary (at $L\to\infty$) where particles are absorbed. In this
case one has indeed $j=\phi_c\equiv j_c$ \cite{Levi05,deMa84}.

For densities not too far above $\rho_c$, there is an initial
metastable regime of finite duration~\cite{Kaup05}.  After this the
system starts to coarsen, i.e., ``small'' condensates at finite
distance start to evolve such that larger condensates grow at the
expense of smaller condensates. In the asymmetric ZRP the mean
distance grows as $\sqrt{t}$ \cite{Gros03a,Godr03}, until eventually
only a single very large condensate remains. This condensate moves on
the lattice on time scales $L^p$ where $p>1$ depends on the details of
the hopping rates \cite{Godr05}.  We conclude that on time scales
$t'=Lt$ the mean separation between condensates is proportional to
$\sqrt{L}$. Hence the local density $\rho(x,t')$, coarse-grained over
length segments proportional to $L$, remains {\it unchanged} on the
Eulerian time scale.  On the other hand, between the condensates the
system has an average background density $\rho_c$ and the current has
its stationary value $j=j_c$ irrespective of $\rho$ \cite{Gros03a}.
Therefore $\partial_x j(\rho)=0$ and (\ref{1-1}) is satisfied even
though the system is not stationary, but coarsens.

Furthermore, one expects that kinematic waves travel with critical
collective velocity $v^\ast_{\text{coll}}\equiv
v_{\text{coll}}(\rho^\ast)$ between condensates and that they are
absorbed into a condensate when they reach it after a time of order
$\sqrt{L}$. Hence the long-time average collective velocity vanishes,
thus allowing for identification of $v_{\text{coll}}$ with
$v_{\text{char}}=\partial_\rho j = 0$ for $\rho > \rho_c$.

It remains to investigate the situation where a finite fraction of the
chain is initially subcritical, whereas some neighbouring domain, also
of length proportional to $L$, is supercritical. Inside each domain
the dynamics are described by the previous considerations. In order to
investigate the boundary between the domains we consider first a
domain boundary characterized by $\rho_k(0) < \rho_c$ for $k<k_0$ and
$\rho_k(0) > \rho_c$ for $k\geq k_0$. We consider two piecewise
constant density profiles with $\rho_\text{left}$
($\rho_\text{right}$) as the density of the subcritical
(supercritical) domain. Without loss of generality we set
$k_0=0$. Hence the two domains are connected by a density jump which
on the macroscopic scale corresponds to a shock.  In the supercritical
domain $k>0$ one expects at Eulerian time scale a series of small
condensates, separated by a fluctuating background with average
density $\rho_c$. We define $k_{\text{left}} = O(\sqrt{L})$ as the
position of the leftmost condensate.  In the whole supercritical
domain the current is $j_c$.

In the subcritical domain ($\rho_\text{left} < \rho_c$) one has
$j<j_c$. Hence the influx into the supercritical domain is less than
the flux inside and as a result the domain boundary moves towards the
first condensate. When it reaches $k_{\text{left}}$ the flux $j$ from
the subcritical region to its left into the condensate becomes smaller
than the flux $j_c$ out of the condensate into the supercritical
region.  As a result the condensate shrinks in size and finally
disappears.  Then the domain boundary moves on until it hits the
second condensate and the process of condensate annihilation sets in
again. Due to mass conservation the domain boundary thus moves into
the supercritical domain and ``eats it up''. The shock separating the
two domains is stable according to the usual stability criterion since
$v_\text{left} > v_s > v_\text{right}$. Notice that, as discussed
above, $v_\text{right}=0$ for $\rho_\text{right}>\rho_c$.

If $\rho_\text{left} = \rho_c$, mass conservation results in a vanishing shock
velocity.  Microscopically one expects a coarsening domain with
background density $\rho_c$ to the right of the critical domain which
has $\rho=\rho_c$, but no condensates.  The microscopic structure of
the shock is not a jump in the background density, instead it
originates from the condensates in the supercritical domain.  The
microscopic position of the shock is determined by the fluctuations of
the position of the leftmost condensate.

Now we consider the space-reflected case where $\rho_\text{left} > \rho_c$ and
$\rho_\text{right} < \rho_c$. In this case the rightmost condensate in the
supercritical region serves as a source with constant flux $j_c$ that
feeds into the subcritical domain with $j<j_c$. Thus one expects a
constant density profile with $\rho=\rho_c$ to the right of the
rightmost condensate up to the beginning of the subcritical domain.
Hence effectively one has a critical region (initially of a size of
the order $\sqrt{L}$) connected to a macroscopic subcritical domain.
The domain boundary moves with collective velocity
$v_{\text{coll}}^\ast$ and to its right a rarefaction wave develops
according to the entropy solution of (\ref{1-1}).  For
$v_{\text{coll}}^\ast=0$ the critical domain with $\rho_c$ does not
grow on the Eulerian time scale.

Therefore, for any domain boundary between subcritical and supercritical segments,
the coarse grained time evolution of the density
profile can be computed from (\ref{1-1}) with the prescription
\bel{1-9} v_{\text{char}} =
\begin{cases}
  \partial_\rho j & \text{for $0 \leq \rho \leq \rho_c$} \\ 0 &
  \text{for $\rho > \rho_c$}.
\end{cases}
\ee The characteristic velocity of the hydrodynamic equation and the
collective velocity of the lattice gas coincide. Notice that for the
TAZRP $j=\phi$ and therefore for $\rho\leq \rho_c$ the collective
velocity is proportional to the inverse of the compressibility
\bel{1-10} 
\kappa =\phi\frac{\partial \rho}{\partial \phi} =
\frac{1}{L}(\exval{N^2}-\exval{N}^2) 
\ee 
of the lattice gas in the grand-canonical ensemble.

\section{Simulation results}
\label{s:sim}
In this section we present the results of Monte Carlo investigations
which confirm the preceding theoretical analysis.  We simulated a
model in which particles hop to the right with rates
\begin{equation}
w_n=1+\frac{b}{n}. \label{e:w}
\end{equation}
In~\cite{Oloa98} it was observed that, for this choice of $w_n$, one
sees condensation for $b>2$.  The critical density $\rho_c=1/(b-2)$
and the critical collective velocity is given by~\cite{Gros03a}
\begin{equation}
v^*_\text{\text{coll}}=
\begin{cases}
0 & \text{for $2<b\leq3$} \\ \frac{(b-3)^2 (b-2)^2}{(b-1)^2} &
\text{for $b>3$}.
\end{cases} \label{e:vcoll}
\end{equation}
Note that the collective velocity vanishes for $2<b\leq3$ because the
compressibility is infinite.

\begin{figure}
\begin{center}
\vspace{-2\baselineskip}
\psfrag{density}[Bc][Tc]{\small{$\rho$}}
\psfrag{k}[Tc][Bc]{\small{$k/10^5$}}
\psfrag{0}[Tc][Tc]{\scriptsize{0.0}}
\psfrag{20000}[Tc][Tc]{\scriptsize{0.2}}
\psfrag{40000}[Tc][Tc]{\scriptsize{0.4}}
\psfrag{60000}[Tc][Tc]{\scriptsize{0.6}}
\psfrag{80000}[Tc][Tc]{\scriptsize{0.8}}
\psfrag{100000}[Tc][Tc]{\scriptsize{1.0}}
\psfrag{120000}[Tc][Tc]{\scriptsize{1.2}}
\psfrag{140000}[Tc][Tc]{\scriptsize{1.4}}
\psfrag{160000}[Tc][Tc]{\scriptsize{1.6}}
\psfrag{180000}[Tc][Tc]{\scriptsize{1.8}}
\psfrag{200000}[Tc][Tc]{\scriptsize{2.0}}
\psfrag{0.0}[Cr][Cr]{\scriptsize{0.0}}
\psfrag{0.5}[Cr][Cr]{\scriptsize{0.5}}
\psfrag{1.0}[Cr][Cr]{\scriptsize{1.0}}
\psfrag{1.5}[Cr][Cr]{\scriptsize{1.5}}
\psfrag{2.0}[Cr][Cr]{\scriptsize{2.0}}
\psfrag{2.5}[Cr][Cr]{\scriptsize{2.5}}
\psfrag{3.0}[Cr][Cr]{\scriptsize{3.0}}
\psfrag{3.5}[Cr][Cr]{\scriptsize{3.5}}
\psfrag{4.0}[Cr][Cr]{\scriptsize{4.0}}
\psfrag{4.5}[Cr][Cr]{\scriptsize{4.5}}
\subfigure[~$t=0.4\times 10^4$, single realization]{\includegraphics*[width=0.47\columnwidth]{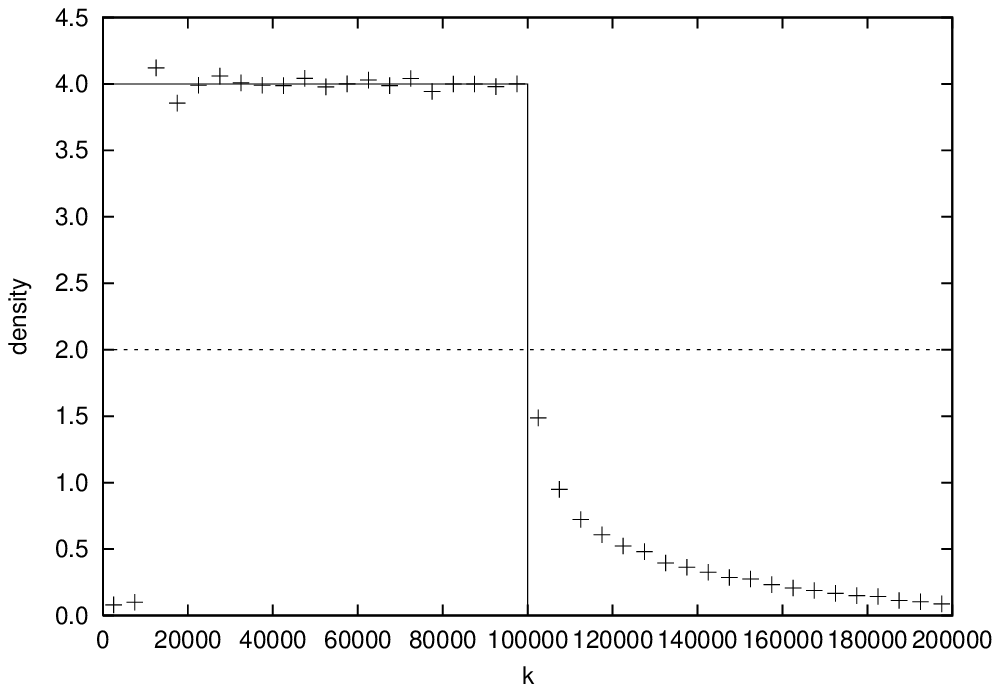}} \qquad
\subfigure[~$t=0.4\times 10^4$, average over histories]{\includegraphics*[width=0.47\columnwidth]{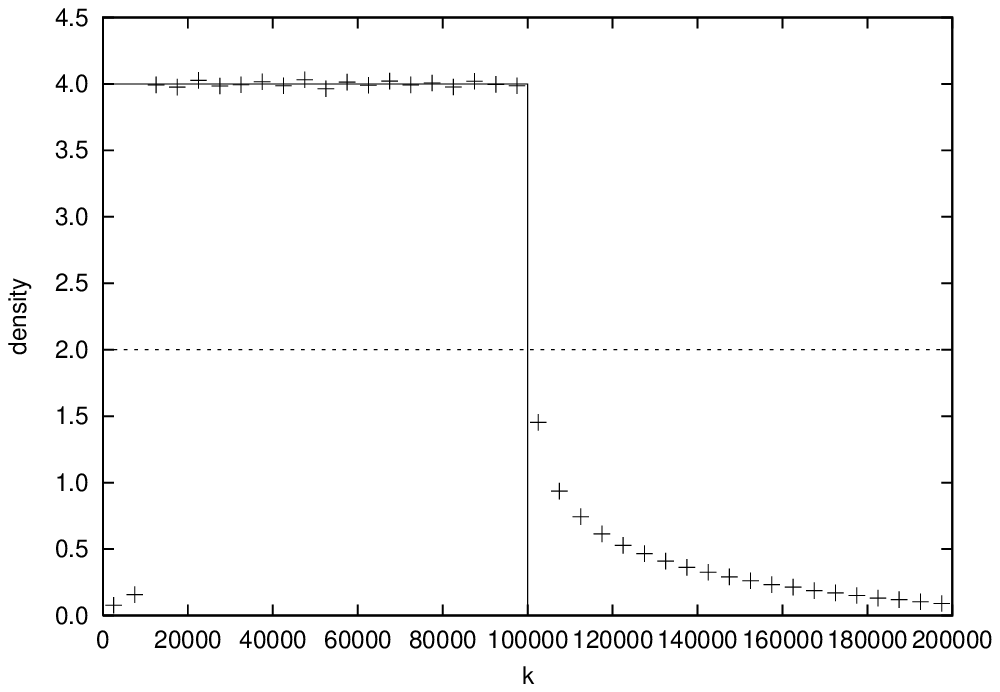}}
\subfigure[~$t=0.8\times 10^4$, single realization]{\includegraphics*[width=0.47\columnwidth]{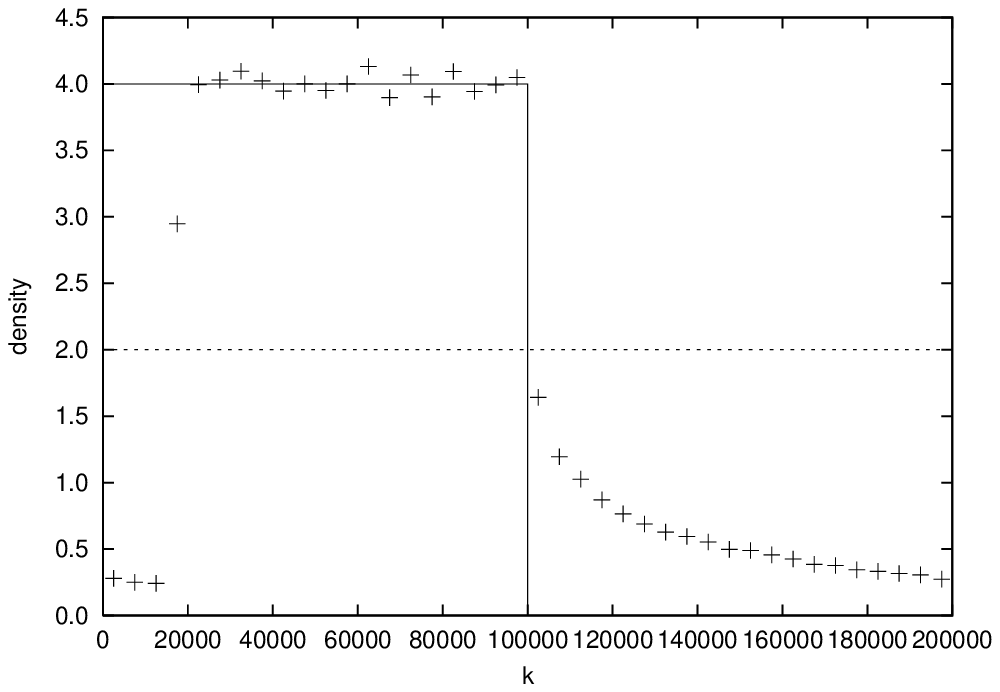}} \qquad
\subfigure[~$t=0.8\times 10^4$, average over histories]{\includegraphics*[width=0.47\columnwidth]{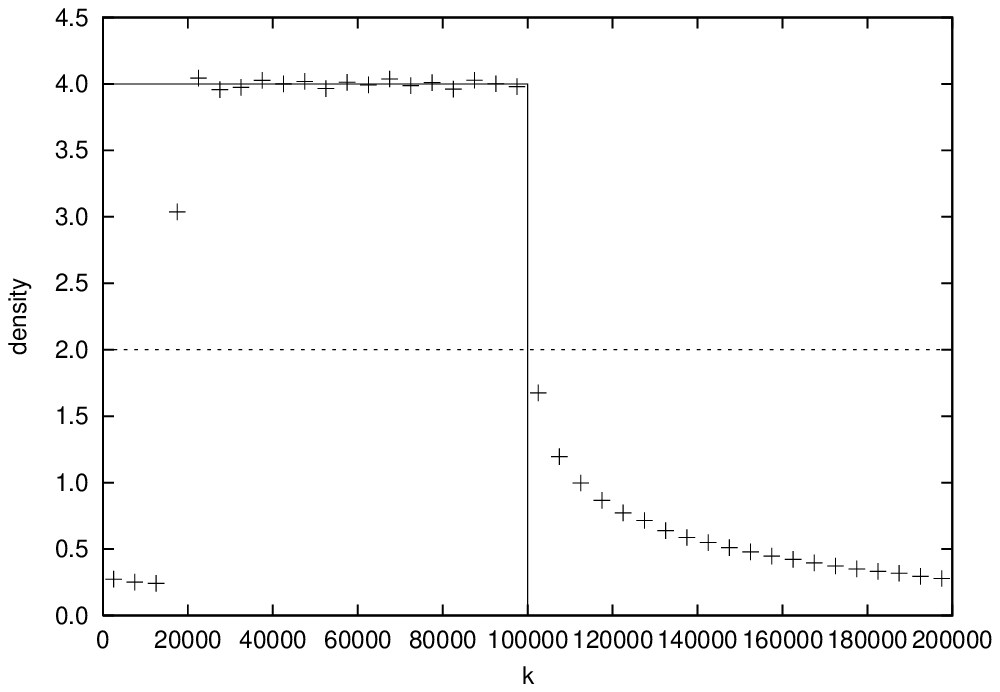}}
\subfigure[~$t=1.6\times 10^4$, single realization]{\includegraphics*[width=0.47\columnwidth]{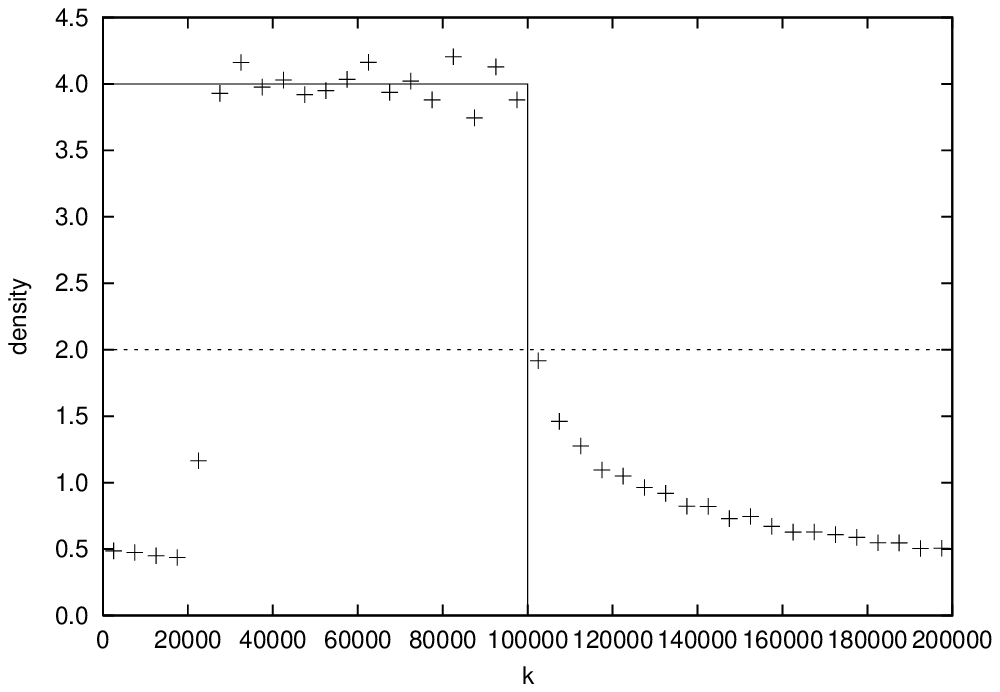}} \qquad
\subfigure[~$t=1.6\times 10^4$, average over histories]{\includegraphics*[width=0.47\columnwidth]{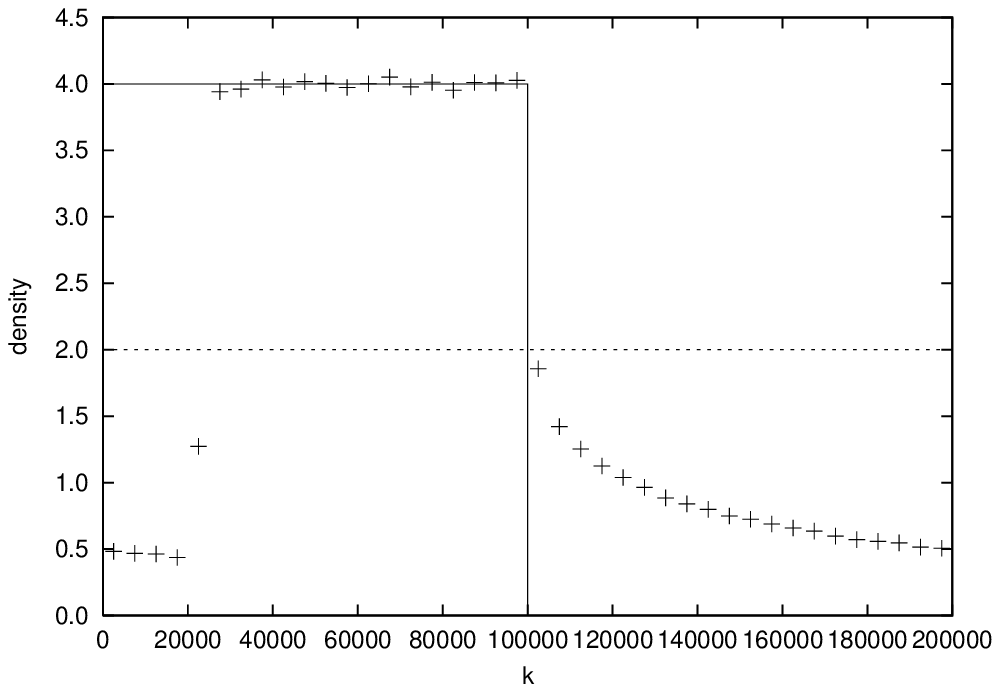}}
\caption{Monte Carlo simulation results for TAZRP on a ring of size
$L=2\times10^5$ where $w_n$ is given by~\eqref{e:w} with $b=2.5$.  The
data points are coarse-grained density profiles, averaged over $5
\times 10^3$ sites, at increasing times for a single realization (left
column) and an average over 10 realizations (right column).  The
initial profile (solid line) and final density (dotted) are shown for
comparison.}
\label{f:b25}
\end{center}
\end{figure}

\begin{figure}
\begin{center}
\vspace{-2\baselineskip}
\psfrag{density}[Bc][Tc]{\small{$\rho$}}
\psfrag{k}[Tc][Bc]{\small{$k/10^5$}}
\psfrag{0}[Tc][Tc]{\scriptsize{0.0}}
\psfrag{20000}[Tc][Tc]{\scriptsize{0.2}}
\psfrag{40000}[Tc][Tc]{\scriptsize{0.4}}
\psfrag{60000}[Tc][Tc]{\scriptsize{0.6}}
\psfrag{80000}[Tc][Tc]{\scriptsize{0.8}}
\psfrag{100000}[Tc][Tc]{\scriptsize{1.0}}
\psfrag{120000}[Tc][Tc]{\scriptsize{1.2}}
\psfrag{140000}[Tc][Tc]{\scriptsize{1.4}}
\psfrag{160000}[Tc][Tc]{\scriptsize{1.6}}
\psfrag{180000}[Tc][Tc]{\scriptsize{1.8}}
\psfrag{200000}[Tc][Tc]{\scriptsize{2.0}}
\psfrag{0.0}[Cr][Cr]{\scriptsize{0.0}}
\psfrag{0.2}[Cr][Cr]{\scriptsize{0.2}}
\psfrag{0.4}[Cr][Cr]{\scriptsize{0.4}}
\psfrag{0.6}[Cr][Cr]{\scriptsize{0.6}}
\psfrag{0.8}[Cr][Cr]{\scriptsize{0.8}}
\psfrag{1.0}[Cr][Cr]{\scriptsize{1.0}}
\psfrag{1.2}[Cr][Cr]{\scriptsize{1.2}}
\psfrag{1.4}[Cr][Cr]{\scriptsize{1.4}}
\subfigure[~$t=0.4\times 10^4$, single realization]{\includegraphics*[width=0.47\columnwidth]{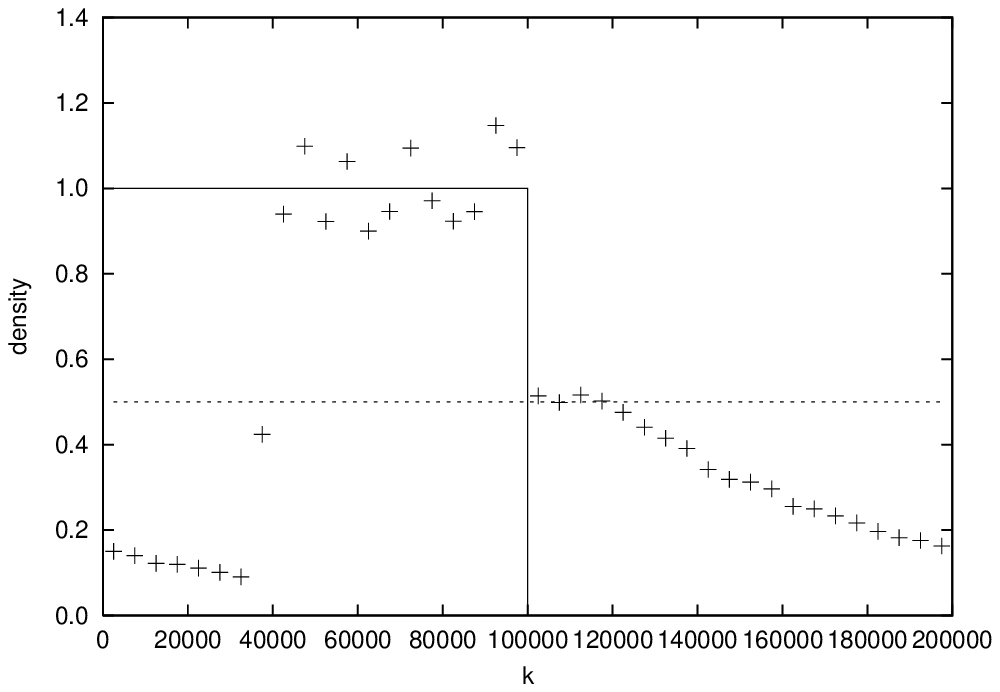}} \qquad
\subfigure[~$t=0.4\times 10^4$, average over histories]{\includegraphics*[width=0.47\columnwidth]{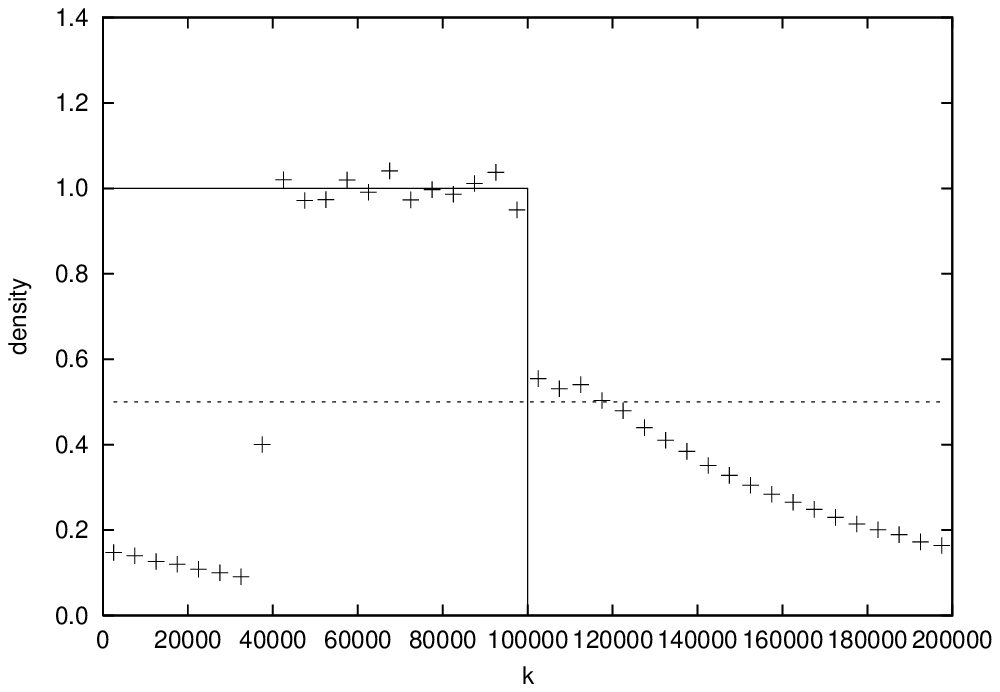}}
\subfigure[~$t=0.8\times 10^4$, single realization]{\includegraphics*[width=0.47\columnwidth]{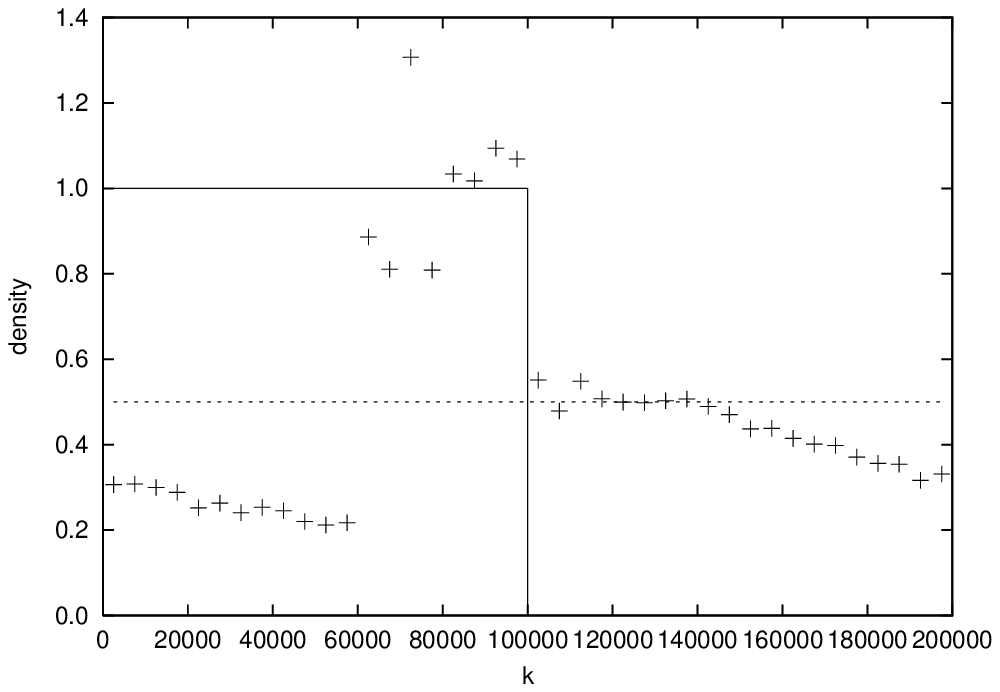}} \qquad
\subfigure[~$t=0.8\times 10^4$, average over histories]{\includegraphics*[width=0.47\columnwidth]{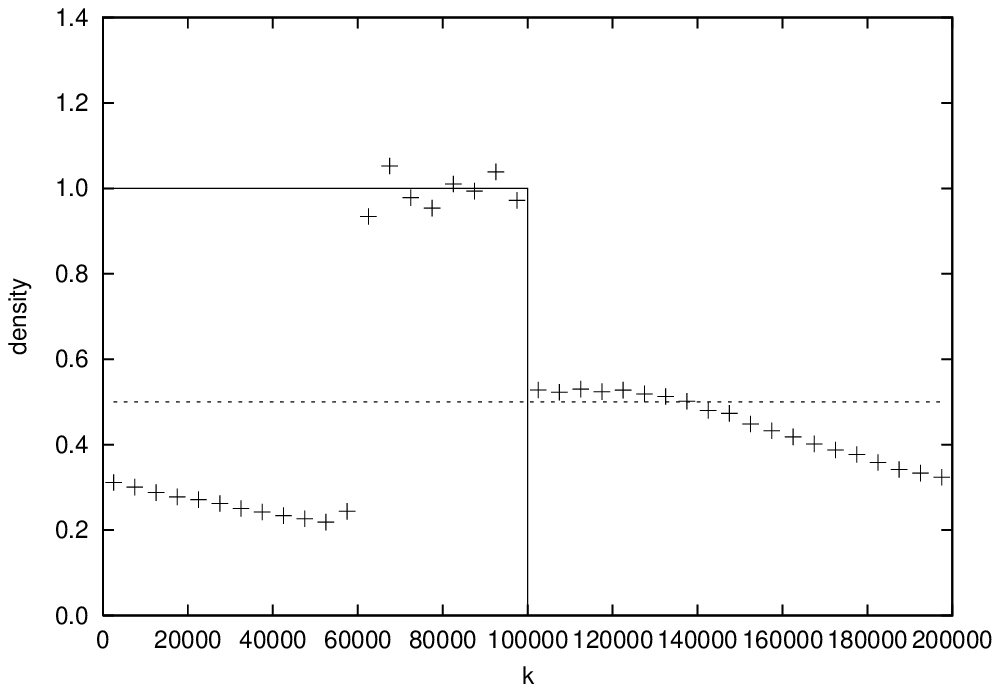}}
\subfigure[~$t=1.6\times 10^4$, single realization]{\includegraphics*[width=0.47\columnwidth]{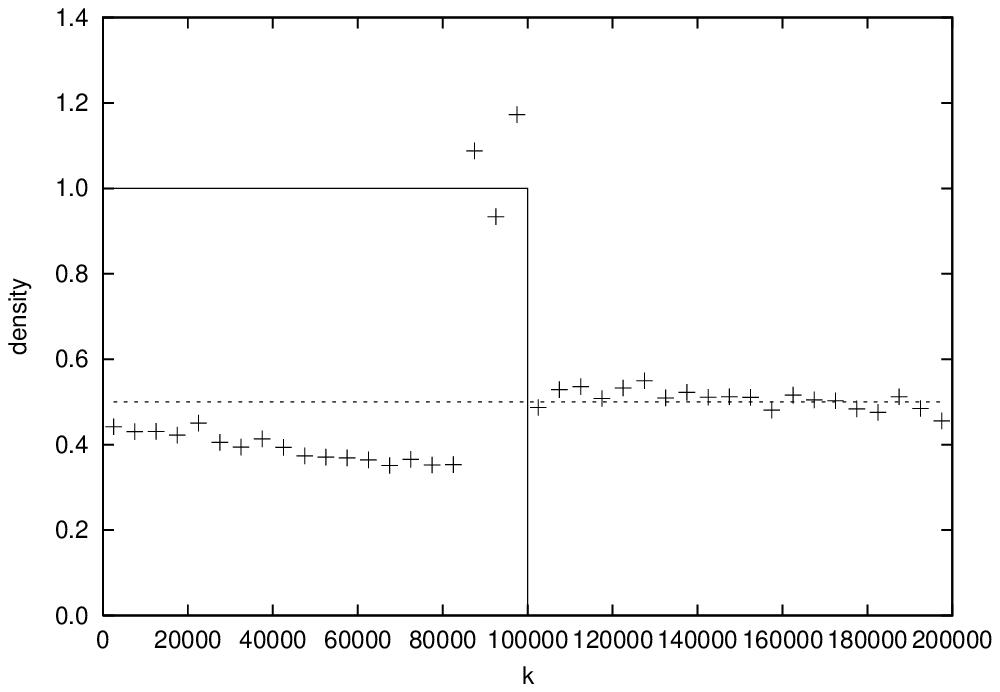}} \qquad
\subfigure[~$t=1.6\times 10^4$, average over histories]{\includegraphics*[width=0.47\columnwidth]{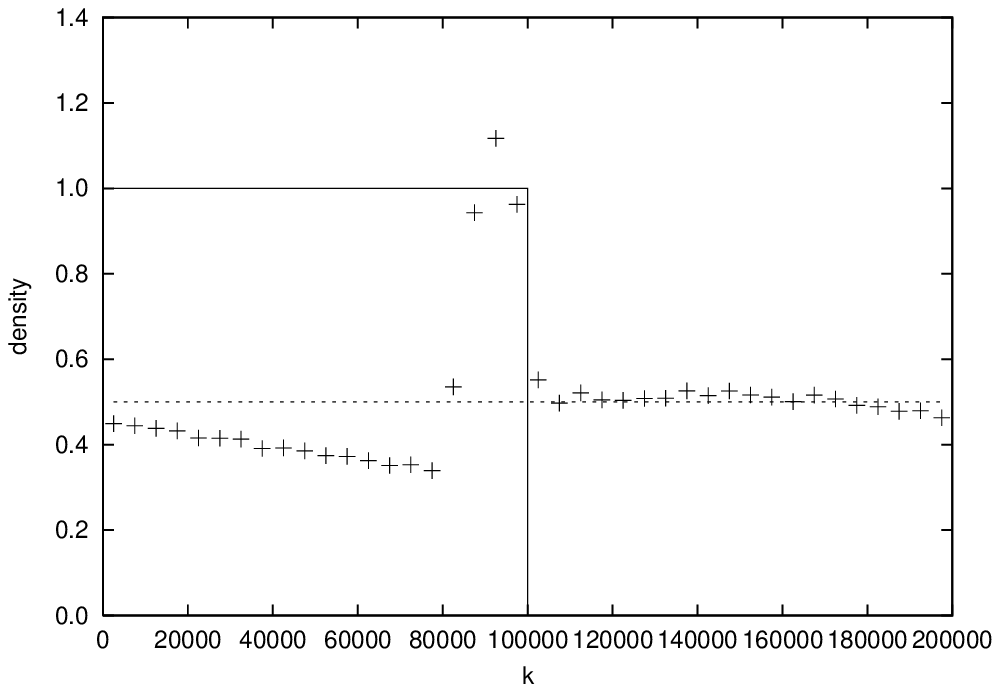}\label{fs:sub}}
\caption{Same as Fig.~\ref{f:b25} but for $b=4$.  From (f) we estimate
the speed of movement for the righthand boundary of the critical
domain (i.e., the flat section with $\rho \approx \rho_c = 0.5$) as $7500/16000 \approx 0.47$, to be compared with the
theoretical prediction, from~\eqref{e:vcoll}, of $4/9 \approx 0.44$.}
\label{f:b4}
\end{center}
\end{figure}

In Figs.~\ref{f:b25} and~\ref{f:b4} we show results for $b=2.5$ and
$b=4$ respectively.
In both cases we simulated a periodic lattice starting from an initial
condition of $\rho_k(0)=2\rho_c$ for $0<k<L/2$ and $\rho_k(0)=0$
elsewhere.  The lefthand columns of the figures contain snapshots of
the coarse-grained density profile for a single realization at
increasing times whereas the righthand columns show an additional
average over stochastic histories.  At the left boundary of the
supercritical region one clearly sees a right-moving shock front which
``eats up'' the condensation regime.  Figure~\ref{f:b4} also clearly
demonstrates the growth of a critical domain ($\rho=\rho_c$) whose
right boundary moves with a speed consistent with~\eqref{e:vcoll}. 

The ensemble average gives qualitatively the same picture as the
coarse-grained space average for a single realization, thus
illustrating the self-averaging nature of the process.  For longer
times, diffusive fluctuations smooth out the the sharp domain
boundaries when the profile is averaged over histories.

\section{Conclusions}
In summary, we have argued that the hydrodynamic description
(\ref{1-1}) of the ZRP remains valid for supercritical densities even
though locally the system is not stationary under Eulerian scaling. In
order give a meaning to (\ref{1-1}) we have analyzed the coarsening
process and found that the continuity equation has to be supplemented
by the results (\ref{1-8}), (\ref{1-9}) for the current.  We have
demonstrated the validity of the theory by Monte Carlo simulation of
the TAZRP.

We expect similar analysis to be valid for the symmetric ZRP, under
diffusive scaling $L\to\infty$ with $k = xL$, $t'=tL^2$.  In this case
one gets the conservation law $\partial_{t'} \rho + \partial_x^2 \phi(\rho)
= 0$. Inside a supercritical domain one has $\phi(\rho)=\phi_c$,
corresponding to a vanishing collective diffusion
coefficient. Adapting the considerations of the coarsening process to
this situation one arrives at a solution that is analogous to the free
boundary solution of the phase-segregation problem in the
low-temperature phase of higher dimensional lattice gases
\cite{Spoh91}.

\acknowledgments
We are grateful for the hospitality of the
Isaac Newton Institute, Cambridge where most of this work was carried
out.  We benefited from many stimulating discussions with other participants in the programme ``Principles of the Dynamics of Nonequilibrium Systems''; in particular, we would like to thank Mustansir Barma and Stefan Grosskinsky.



\begin{thebibliography}{99}

\bibitem{Spit70} F. Spitzer,
Interaction of Markov Processes,
Adv. Math. {\bf 5} 246--290 (1970).

\bibitem{Andj82} E.D. Andjel, 
Invariant measures for the zero range process,
Ann. Probab. {\bf 10} 525 (1982).

\bibitem{Reza91} F. Rezakhanlou, Hydrodynamic limit for attractive particle
systems on  $\mathbb{Z}^d$, Comm. Math. Phys. {\bf 140} 417--448 (1991).

\bibitem{Kipn99}
C. Kipnis and C. Landim, {\it Scaling limits of interacting
particle systems} (Springer, Berlin, 1999).

\bibitem{Evan05} M.R. Evans and T. Hanney, 
Nonequilibrium statistical mechanics of the zero-range process and 
related models,
J. Phys. A: Math. Gen. {\bf 38} R195--R240 (2005).

\bibitem{Oloa98} O.J. O'Loan, M.R. Evans, and M.E. Cates,
Jamming transition in a homogeneous one-dimensional system: The bus route model,
Phys. Rev. E {\bf 58} 1404--1418 (1998).

\bibitem{Jeon00a} I. Jeon and P. March, 
Condensation transition for zero-range invariant measures, 
Can. Math. Soc. Conf. Proc {\bf 26} 233--244 (2000).

\bibitem{Jeon00b} I. Jeon and P. March,
Size of the largest cluster under zero-range invariant measures,
Ann. Probab. {\bf 28} 1162--1194 (2000).

\bibitem{Gros03a}
S. Grosskinsky, G.M. Sch\"utz, and H. Spohn,
Condensation in the zero range process: stationary and dynamical properties
J. Stat. Phys. {\bf 113} 389--410 (2003).

\bibitem{Godr03}
C. Godr\`eche, 
Dynamics of condensation in zero-range processes,
J. Phys. A: Math. Gen. {\bf 36} 6313--6328 (2003).

\bibitem{Kaup05} J. Kaupu{\v z}s, R. Mahnke, and R.J. Harris,
Zero-range model of traffic flow, 
Phys. Rev. E {\bf 72} 056125 (2005). 

\bibitem{Anta00}
T. Antal and G.M. Sch\"utz,
Asymmetric exclusion process with next-nearest-neighbor interaction: Some comments on traffic flow and a nonequilibrium reentrance transition,
Phys. Rev. E {\bf 62} 83--93 (2000).

\bibitem{Kafr02} Y. Kafri, E. Levine, D. Mukamel, G.M. Sch\"{u}tz, and
J. T\"or\"ok, 
Criterion for phase separation in one-dimensional driven systems,
Phys. Rev. Lett. {\bf 89} 035702 (2002)

\bibitem{Toth03} B. T\'oth and B. Valk\'o,
Onsager relations and Eulerian hydrodynamic limit for systems with several conservation laws,
J. Stat. Phys. {\bf 112} 497--521 (2003).

\bibitem{Schu03} G.M. Sch\"utz,
Critical phenomena and universal dynamics in one-dimensional driven diffusive systems with two species of particles,
J. Phys. A.: Math. Gen, {\bf 36} R339--R379 (2003).

\bibitem{Schu97} 
G.M. Sch\"utz, 
The Heisenberg chain as a dynamical model for protein synthesis---Some theoretical and experimental results,
Int. J. Mod. Phys. B {\bf 11} 197--202 (1997).

\bibitem{Nish05}
K. Nishinari, Y. Okada, A. Schadschneider, and D. Chowdhury, 
Intracellular transport of single-headed molecular motors KIF1A,
Phys. Rev. Lett. {\bf 95} 118101 (2005).

\bibitem{Lipo02} A. Lipowski and M. Droz,
Urn model of separation of sand, 
Phys. Rev. E {\bf 65} 031307 (2002).

\bibitem{Shim04}
G.M. Shim, B.Y. Park, J.D. Noh, and H. Lee,
Analytic study of the three-urn model for separation of sand,
Phys. Rev. E {\bf 70} 031305 (2004).

\bibitem{Schu96}
G.M. Sch\"utz, R. Ramaswamy, and M. Barma,
Pairwise balance and invariant measures for generalized exclusion processes,
J. Phys. A: Math. Gen. {\bf 29} 837--843 (1996).

\bibitem{Spoh91}
H. Spohn, {\it Large Scale dynamics of interacting particle systems},
(Springer, Berlin, 1991)

\bibitem{Schu00} G.M. Sch\"utz,
{\it Exactly solvable models for many-body systems far from equilibrium},
in {\it Phase
Transitions and Critical Phenomena} vol 19, ed. C. Domb and J.
Lebowitz (London: Academic, 2001), pp 1--251.

\bibitem{Ligh55} M.J. Lighthill and G.B. Whitham,
On Kinematic Waves II:  A Theory of Traffic Flow on Long Crowded Roads,
Proceedings of the  Royal Society of  London Series A
{\bf  229} 317--345 (1955).

\bibitem{deMa84}
A. De Masi and P. Ferrari, A remark on the hydrodynamics of the zero-range processes,
J. Stat. Phys. {\bf 36} 81--87 (1984).

\bibitem{Schu93}
G. Sch\"utz and E. Domany,
Phase transitions in an exactly soluble one-dimensional exclusion process,
J. Stat. Phys. {\bf 72} 277--296 (1993).

\bibitem{Derr93}
B. Derrida, V. Hakim, M.R. Evans, and V. Pasquier,
Exact solution of a 1D asymmetric exclusion model using a matrix formulation, 
J. Phys. A: Math. Gen. {\bf 26} 1493--1517 (1993).

\bibitem{Kolo98}
A.B. Kolomeisky, G.M. Sch\"{u}tz, E.B. Kolomeisky and J.P. Straley,
Phase diagram of one-dimensional driven lattice gases with open boundaries,
J. Phys. A: Math. Gen. {\bf 31} 6911--6919 (1998).

\bibitem{Gros03b}
S. Grosskinsky and H. Spohn,
Stationary measures and hydrodynamics of zero range processes with several species of particles,
Bull. Braz. Math. Soc. New Series 34, 489--507 (2003).

\bibitem{Baha04} C. Bahadoran,
Hydrodynamics of asymmetric particle systems with open boundaries, 
Oberwolfach Report 43/2004, 2290--2292.

\bibitem{Levi05} E. Levine, D. Mukamel, and G.M. Sch\"utz, 
Zero-range process with open boundaries,
J. Stat. Phys. {\bf 120} 759--778 (2005).

\bibitem{Godr05} C. Godr\`eche and J.M. Luck,
Dynamics of the condensate in zero-range processes 
J. Phys. A: Math. Gen. {\bf 38} 7215--7237 (2005).

\end{thebibliography}
\end{document}